\documentclass{ws-ijmpd-arXiv11081995}

\usepackage{amsmath,amssymb,revsymb,graphicx,dcolumn}

\newcommand{\beq}{\begin{equation}}
\newcommand{\eeq}{\end{equation}}
\newcommand{\beqa}{\begin{eqnarray}}
\newcommand{\eeqa}{\end{eqnarray}}
\newcommand{\bsubeqs}{\begin{subequations}}
\newcommand{\esubeqs}{\end{subequations}}
\newcommand{\half}{{\textstyle\frac{1}{2}}}

\begin{document}

\markboth{V. Emelyanov, F.R. Klinkhamer}
{Vector-field model with compensated $\Lambda$ and FRW phase}

%
\catchline{}{}{}{}{}
%

\title{\mbox{VECTOR-FIELD MODEL WITH COMPENSATED COSMOLOGICAL}
       \mbox{CONSTANT AND RADIATION-DOMINATED FRW PHASE}}

\author{V. EMELYANOV}
\address{Institute for Theoretical Physics, University of Karlsruhe,\\
Karlsruhe Institute of Technology, 76128 Karlsruhe, Germany\\
slawa@particle.uni-karlsruhe.de}

\author{F.R. KLINKHAMER}
\address{Institute for Theoretical Physics, University of Karlsruhe,\\
Karlsruhe Institute of Technology, 76128 Karlsruhe, Germany\\
frans.klinkhamer@kit.edu}

\maketitle


\begin{abstract}
A special model of a massless vector-field is presented,
which has an extra modified-gravity-type interaction term in the action.
The cosmology of the model is studied
with standard noninteracting relativistic matter added.
It is found that this
cosmology can have an early phase where the vector-field
starts to compensate a (Planck-scale) cosmological constant and a late
Friedmann--Robertson--Walker (FRW) phase where the relativistic-matter
energy density dominates the dynamic vacuum energy density.
\vspace*{1\baselineskip}\newline
\textit{Journal}: Int. J. Mod. Phys. D 21 (2012) 1250025
\vspace*{.75\baselineskip}\newline
\textit{Preprint}: arXiv:1108.1995 
\end{abstract}

\newpage
\section{Introduction}\label{sec:intro}

The cosmological constant
problem\cite{CCP-review-Weinberg,CCP-review-SahniStarobinsky}
is perhaps the most important
outstanding question
of modern physics. In a nutshell, the problem is to explain why
the energy scale corresponding
to the measured value of the cosmological constant $\Lambda$
is negligible compared to the energy
scales of elementary particle physics,
$|\Lambda|^{1/4}\ll E_\text{QCD}\ll E_\text{ew}\ll
E_\text{Planck} \sim 10^{18}\;\text{GeV}$.

The cosmological constant $\Lambda$ can be
canceled dynamically and without fine-tuning in
a particular massless-vector-field model.\cite{Dolgov1985,Dolgov1997}
But this model runs into two obstacles.
First, the local Newtonian dynamics is ruined.\cite{RubakovTinyakov1999}
Second, the final phase of the model universe expands so rapidly
(scale factor $a \propto t$)
that any standard matter contribution
($\rho_M \propto 1/a^p$ with $p=4$ for relativistic matter
or $p=3$ for nonrelativistic matter)
becomes irrelevant compared to
the decreasing remnant vacuum energy density
($\rho_V \propto 1/t^2$).

Following previous work on the $q$--theory
approach\cite{KV2008a,KV2008b,KV2010,KV2009-electroweak,K2010-electroweak,K2011-electroweak,KV2011-review}
to the main cosmological constant problem, an extended model with two
massless vector-fields has been
found,\cite{EmelyanovKlinkhamer2011-No1,Klinkhamer2011}
which evades the first obstacle regarding the local Newtonian dynamics.
Here, we present another model with a single vector-field
and a modified-gravity-type interaction term, which
circumvents the second obstacle by having a
final phase with slower expansion ($a \propto t^{1/2}$).

\section{Model and Ansatz}
\label{sec:Model-and-Ansatz}

Consider the following model of a massless vector-field
$A_{\alpha}(x)$ with an effective action ($\hbar=c=1$)
\bsubeqs\label{eq:model-action-Q3-general}
\begin{equation}\label{eq:model-action}
S_\text{eff} =
-\int\,d^4x\,\sqrt{-g}\,\left[ \frac{R}{16\pi\,G} + \epsilon(Q_3)
+ \Lambda + f(R)\,A_{\alpha}A^{\alpha}
+ \mathcal{L}_M
\right],
\end{equation}
where $R$ is the Ricci scalar,
$\Lambda$ the effective cosmological constant,
and $f(R)$ a function to be specified later.
For the moment, the contribution of other matter fields
is set to zero, $\mathcal{L}_M=0$.
The $q$--theory-type scalar\cite{KV2008a,KV2008b,KV2010}
entering the above action is defined as follows:
\beqa\label{eq:Q3-general}
Q_3(x) \equiv g^{\alpha\beta}(x)\,A_{\alpha;\beta}(x)\,,
\eeqa
\esubeqs
with the spacetime coordinate $(x)=(x^{1},\, x^{2},\, x^{3},\, t)$
as argument and the semicolon standing for covariant derivation.
The particular contraction \eqref{eq:Q3-general} was mentioned
in Ref.~\refcite{Dolgov1997} but not studied in detail.

In the following, we consider a
spatially-flat Friedmann--Robertson--Walker (FRW)
universe with scale factor $a(t)$ for cosmic time $t$ and
take the isotropic \textit{Ansatz}\cite{Dolgov1985,Dolgov1997}
for the vector-field:
\bsubeqs\label{eq:Ansaetze-metric-vector}
\beqa\label{eq:Ansatz-metric}
\Big(g_{\alpha\beta}(t) \Big) &=&
\Big(\textrm{diag}\big[1,\,-a^2(t),\,-a^2(t),\,-a^2(t)\big]\Big)\,,
\\[3mm]\label{eq:Ansatz-vector}
A_{\alpha}(t) &=& A_{0}(t)\;\delta_{\alpha}^0\,.
\eeqa
\esubeqs
The \textit{Ans\"{a}tze} \eqref{eq:Ansaetze-metric-vector} give
\bsubeqs\label{eq:Ansaetze-R-Q3}
\beqa\label{eq:Ansatz-R}
R &=& -6\,(\dot{H}+2\, H^2)\,,
\\[2mm]
\label{eq:Ansatz-Q3}
Q_3 &=& \dot{A}_0 + 3\,H\,A_{0}\,,
\eeqa
\esubeqs
where the overdot stands for differentiation with respect to the
cosmic time $t$ and the Hubble parameter $H(t)$ is defined by
$H(t)\equiv \dot{a}(t)/a(t)$.

Note that the variable $(Q_3)^2=(\dot{A}_0 + 3\,H\,A_{0})^2$
differs from the
variable $(Q_1)^2 \equiv A_{\alpha;\beta}\,A^{\alpha;\beta}
=\dot{A}_0^2 + 3\,H^2\,A_{0}^2$
discussed in Refs.~\refcite{KV2010} and \refcite{EmelyanovKlinkhamer2011-No1}.
For the statics of $q$--theory, the precise realization of the
$q$-type vacuum variable is irrelevant [here, either $Q_3$ or $Q_1$].
But for the dynamics, the realization matters
[here, giving a FRW universe with
$t H(t)$ approaching either the value $1/2$ or the value $1$].
See App.~A of Ref.~\refcite{KV2011-review} for a brief introduction
to the basic ideas of $q$--theory. This introduction to $q$--theory
may give the reader some background information
but is not needed to understand the present article,
which is entirely self-contained.

\section{Reduced Field Equations}
\label{sec:Reduced-Field-Equations}
\subsection{Vector-field equation}
\label{subsec:Vector-field-equation}

The variational principle applied to
the vector-field of action \eqref{eq:model-action}
gives the following equation:
\beqa\label{eq:vector-field-eq-f}
\nabla_{\alpha}\left(\frac{d\epsilon}{dQ_3}\right) &=&
2\, f(R)\, A_{\alpha}\,.
\eeqa
The \textit{Ans\"{a}tze} \eqref{eq:Ansaetze-metric-vector} reduce
this field equation to a single ordinary differential equation (ODE),%
\beqa\label{eq:vector-field-eq-f-ODE}
\zeta\, \ddot{A}_0 + (3\, H\, \zeta + \dot{\zeta})\, \dot{A}_0
+\Big[3\, \dot{H}\, \zeta + 3\, H\, \dot{\zeta} - f(R)\, \Big]\, A_{0}
&=& 0\,,
\eeqa
with $\zeta \equiv (2\,Q_3)^{-1}\,d\epsilon/d Q_3$.

\subsection{Generalized FRW equations}
\label{subsec:Generalized -FRW-equations}

The energy-momentum tensor of the vector-field $A_{\alpha}$ follows
from the variation of the action with respect to
the metric field $g_{\alpha\beta}(x)$. The result is
\beqa\label{eq:energy-mom-tensor}
\hspace*{-4mm}&&
T_{\alpha\beta} = \left[\epsilon - Q_3\,\frac{d\epsilon}{dQ_3} -
A^{\lambda}\nabla_{\lambda}\left(\frac{d\epsilon}{dQ_3}\right)\right]\,
g_{\alpha\beta}
+ 2\,A_{\left(\alpha\right.}\nabla_{\left.\beta\right)}\,
\left(\frac{d\epsilon}{dQ_3}\right)
\nonumber\\[1mm]
\hspace*{-4mm}&&
-2\, \bigg(A^2\, f^{\prime}\, R_{\alpha\beta}-\frac{1}{2}\, A^2\, f\, g_{\alpha\beta}
+ f\, A_{\alpha}A_{\beta} -
\nabla_{\alpha}\nabla_{\beta}\, (f^{\prime}\, A^2)
+ g_{\alpha\beta}\, \nabla^2\big(f^{\prime}\, A^2\big)\bigg),
\eeqa
where the round brackets around spacetime indices denote
symmetrization and the prime on $f$
stands for differentiation with respect to $R$.
With the vector-field equation \eqref{eq:vector-field-eq-f},
the energy-momentum tensor \eqref{eq:energy-mom-tensor} becomes
\beqa\label{eq:energy-mom-tensor-vector-Ansatz}
\hspace*{-10mm}&&
T_{\alpha\beta} = \bigg(\epsilon - Q_3\,\frac{d\epsilon}{dQ_3}\bigg)\,
g_{\alpha\beta}
\nonumber\\[1mm]\hspace*{-10mm}&&
-2\, \bigg(A^2\, f^{\prime}\, R_{\alpha\beta}
+ \frac{1}{2}\, A^2\, f\, g_{\alpha\beta} - f\, A_{\alpha}A_{\beta} -
\nabla_{\alpha}\nabla_{\beta}\, \big(f^{\prime}\, A^2\big)
+ g_{\alpha\beta}\, \nabla^2\, \big(f^{\prime}\, A^2\big)\bigg)\,.
\eeqa

From the gravitational field equations,
the \textit{Ans\"{a}tze} \eqref{eq:Ansaetze-metric-vector}, and the
energy-momentum tensor \eqref{eq:energy-mom-tensor-vector-Ansatz},
the generalized  FRW equations are
\bsubeqs\label{eq:Friedmann-Einstein-FRWeqs}
\beqa\label{eq:Friedmann-FRWeqs}
3\, H^2 &=&(8\pi\,G)\, \big[\Lambda + \rho(A)+ \rho_M\big],
\\[2mm]\label{eq:Einstein-FRWeqs}
2\dot{H} + 3H^2 &=&
(8\pi\,G)\, \big[\Lambda - P(A)- w_M\,\rho_M\big],
\eeqa
\esubeqs
having added a further matter contribution from a
nonzero term $\mathcal{L}_M$ in the action \eqref{eq:model-action},
with $w_M\equiv P_M/\rho_M$ the constant equation-of-state (EOS)
parameter of the corresponding homogeneous perfect fluid.
The vector-field contributions on the right-hand sides of
\eqref{eq:Friedmann-Einstein-FRWeqs} are given by
\bsubeqs\label{eq:rhoA-PA-FRW}
\beqa
\hspace*{-4mm}
\rho(A) &=& +\epsilon - Q_3\,\frac{d\epsilon}{dQ_3}
+ 2\, \bigg[3\,(\dot{H}+H^2)\, f^{\prime}\, A_{0}^2+\frac{1}{2}\,f\, A_{0}^2
- 3\, H\,\frac{d}{dt}\Big(f^{\prime}\, A_{0}^2\Big)\bigg],
\\[3mm]
\hspace*{-4mm}
P(A) &=& - \epsilon + Q_3\,\frac{d\epsilon}{dQ_3}
\nonumber\\[2mm]
\hspace*{-4mm}&&
- 2\, \bigg[(\dot{H} + 3\, H^2)\, f^{\prime}\, A_{0}^2
- \frac{1}{2}\, f\, A_{0}^2
- \frac{d^2}{dt^2}\, \Big(f^{\prime}\, A_{0}^2\Big)
- 2\, H\,\frac{d}{dt}\, \Big(f^{\prime}\, A_{0}^2\Big)\bigg].
\eeqa
\esubeqs

\section{Special Vector-Field Model}
\label{sec:Special-Model}

\subsection{Linear $f(R)$ and quadratic $\epsilon(Q_3)$}
\label{subsec:Linear-f-quadratic-epsilon}

The special model considered in this article has a linear
function $f(R)$ in the interaction term of the
action \eqref{eq:model-action},
\bsubeqs\label{eq:f-epsilon-Ansaetze}
\beq\label{eq:f-Ansatz}
\overline{f}(R) = \kappa\:R\,,
\eeq
and a quadratic vector-field function $\epsilon(Q_3)$,
\beq\label{eq:epsilon-Ansatz}
\overline{\epsilon}(Q_3) = \zeta_{0}\:Q_3^2\,,
\eeq
for nonzero dimensionless coefficients
$\kappa$ and $\zeta_{0}$.
In order to be able to cancel a nonzero cosmological
constant, these two coefficients must obey the condition
\beq\label{eq:zeta-kappa-condition-Ansatz}
\text{sgn}\big(6\,\kappa+5\,\zeta_{0}\big) =
\text{sgn}\big(\Lambda\big)\,,
\eeq
\esubeqs
in terms of the sign function $\text{sgn}(x)\equiv x/|x|$
for $x \ne 0$ and $\text{sgn}(0)\equiv 0$.
The particular form of condition
\eqref{eq:zeta-kappa-condition-Ansatz} will be derived
at the end of this subsection.\footnote{Condition
\eqref{eq:zeta-kappa-condition-Ansatz} is purely technical
and appears because of the simple $\epsilon$--function
used in \eqref{eq:epsilon-Ansatz}. For an appropriate, more complicated,
$\epsilon$--function
(cf. Refs.~\refcite{KV2008a,KV2008b,KV2010,EmelyanovKlinkhamer2011-No1}),
the cancelation of the cosmological constant $\Lambda$ can be expected
to hold for any sign of $\Lambda$.}

The \textit{Ansatz} \eqref{eq:f-Ansatz} reduces
the energy density and the isotropic pressure from \eqref{eq:rhoA-PA-FRW}
to the following expressions:%
\bsubeqs\label{eq:rhoA-PA-Ansatz}
\beqa
\rho(A) &=& +\epsilon - Q_3\,\frac{d\epsilon}{dQ_3}
- 6\,\kappa\,\bigg(H^2\,A_{0}^2 + H\,\frac{d}{dt}\,A_{0}^2\bigg),
\\[1mm]
P(A) &=& -\epsilon + Q_3\,\frac{d\epsilon}{dQ_3}
- 2\,\kappa\,\bigg((4\,\dot{H} + 9\,H^2)\,A_{0}^2
- \frac{d^2}{dt^2}\,A_{0}^2 - 2\,H\,\frac{d}{dt}\,A_{0}^2\bigg).
\eeqa
\esubeqs
Similarly, the \textit{Ansatz} \eqref{eq:epsilon-Ansatz}
reduces the ODE \eqref{eq:vector-field-eq-f-ODE}
to the following equation:
\beq\label{eq:vector-field-eq-special-f}
\ddot{A}_0 + 3\, H\, \dot{A}_0 +
\bigg[3\, \dot{H}- \frac{1}{\zeta_{0}}\, f(R)\bigg]\, A_{0} = 0\,.
\eeq
Inserting the function \eqref{eq:f-Ansatz} in the last ODE
and using \eqref{eq:Ansatz-R} gives
\beq\label{eq:vector-field-eq-special}
\ddot{A}_0 + 3\, H\, \dot{A}_0 + 3\, \dot{H}\, A_{0} +
6\,\frac{\kappa}{\zeta_{0}}\,\big(\dot{H}+2\, H^2\big)\, A_{0} = 0\,.
\eeq
\noindent Equation~\eqref{eq:vector-field-eq-special}
is the core result of this article, as will become
clear in Sec.~\ref{subsec:Heuristic-argument}.

For $t\to\infty$, the asymptotic
solution of \eqref{eq:vector-field-eq-special} is given by
\beq\label{eq:A-H-aymptotic}
A_\text{asymp}(t) = \overline{\theta}\:t\,,\quad
H_\text{asymp}(t) = (1/2)\:t^{-1}\,,
\eeq
with a constant $\overline{\theta}$
to be determined by the solution of the
combined field equations (see Sec.~\ref{subsec:Dimensionless-ODEs}).
Substituting this asymptotic solution into the
energy density $\rho(A)$ and the pressure $P(A)$ from
\eqref{eq:rhoA-PA-Ansatz}, we obtain
\bsubeqs\label{eq:rho-P-aymptotic}
\beqa\label{eq:rho-aymptotic}
\rho_{\Lambda}(A_\text{asymp}) &=&
+\Lambda+
\left[\overline{\epsilon}
- Q_3\,\frac{d\,\overline{\epsilon}}{d Q_3}\right]_{Q_3=5\,\overline{\theta}/2}
-\frac{15}{2}\;\kappa\;\overline{\theta}^{\,2}\,,
\\[1mm]
\label{eq:P-aymptotic}
P_{\Lambda}(A_\text{asymp}) &=&
-\Lambda-
\left[\overline{\epsilon}
- Q_3\,\frac{d\,\overline{\epsilon}}{d Q_3}\right]_{Q_3=5\,\overline{\theta}/2}
+\frac{15}{2}\;\kappa\;\overline{\theta}^{\,2}\,,
\eeqa
\esubeqs
having added the contributions from the cosmological constant
$\Lambda$ in the action \eqref{eq:model-action}. Observe that
$\rho_{\Lambda}(A_\text{asymp}) + P_{\Lambda}(A_\text{asymp}) = 0$.
In fact, \eqref{eq:rho-P-aymptotic} has the particular
structure of $q$--theory.\cite{KV2008a,KV2008b,KV2010}
It is even possible to define a new effective
$\epsilon$--function by making a $Q_3$--independent shift,
$\overline{\epsilon}_\text{eff}(Q_3)=\overline{\epsilon}(Q_3)-
(15/2)\;\kappa\;\overline{\theta}^{\,2}$.

Equation \eqref{eq:rho-P-aymptotic} also explains
condition \eqref{eq:zeta-kappa-condition-Ansatz}:
for the simple quadratic function \eqref{eq:epsilon-Ansatz},
it is possible to cancel $\Lambda$ with an appropriate value
of the linear $\theta$--coefficient from \eqref{eq:A-H-aymptotic}
only if condition \eqref{eq:zeta-kappa-condition-Ansatz} is satisfied.
As mentioned before, the actual value $\overline{\theta}$
follows from the solution of the combined field equations.
Remark, finally, that the expression on the right-hand side of
\eqref{eq:rho-aymptotic} may be called the `dressed' cosmological constant
if $\Lambda$ from the action \eqref{eq:model-action}
is considered to be the `bare' cosmological constant.

\subsection{Heuristic argument}
\label{subsec:Heuristic-argument}

Returning to the vector-field ODE \eqref{eq:vector-field-eq-special},
we can give the following heuristic argument for the appearance of the
FRW-like asymptotic solution \eqref{eq:A-H-aymptotic}, even with a nonzero
cosmological constant $\Lambda$ present in
the action \eqref{eq:model-action}.
The physically relevant case has both coefficients
$\kappa$ and $\zeta_{0}$ nonvanishing, but, for completeness,
also the other cases will be briefly mentioned.

For $\kappa\ne 0$, $\zeta_{0}\ne 0$, and a linear time-dependence
of the vector component $A_{0}(t) \propto t$,
the $H^2 A_{0}$ term in \eqref{eq:vector-field-eq-special}
excludes having $H(t)=\text{const}$.
Purely for dimensional reasons, assume
$H(t)=\widehat{\gamma}\;t^{-1}$
with a numerical constant $\widehat{\gamma}>0$
[excluding the case of a static or contracting Universe
with $\widehat{\gamma} \leq 0\,$]. The first three terms
of the left-hand side of \eqref{eq:vector-field-eq-special}
then cancel by themselves. This leaves the $\kappa$ term
in \eqref{eq:vector-field-eq-special}, which
vanishes for $\widehat{\gamma}=1/2$ corresponding to a
radiation-dominated FRW universe.
Indeed, the functions \eqref{eq:A-H-aymptotic} allow for
a compensation of the
cosmological constant $\Lambda$, as shown by
\eqref{eq:rho-P-aymptotic} and the final ODEs to be
presented in Sec.~\ref{subsec:Dimensionless-ODEs}.

For $\kappa=0$ and $\zeta_{0}\ne 0$, the ODE \eqref{eq:vector-field-eq-special}
also has the de-Sitter solution with constant $A_{0}$ and $H$.
As said before, precisely this solution is forbidden by having the
$\kappa\,H^2 A_{0}$ term in \eqref{eq:vector-field-eq-special}
if $\kappa\ne 0$.

For $\kappa\ne 0$ and $\zeta_{0}=0$,
\eqref{eq:vector-field-eq-special} implies $A_{0}=0$
(i.e., $\overline{\theta}=0$) for generic $H(t)$ and the
cosmological constant can no longer be canceled,
as exemplified by \eqref{eq:rho-P-aymptotic}.

For $\kappa=\zeta_{0}=0$
in the \textit{Ans\"{a}tze} \eqref{eq:f-epsilon-Ansaetze},
the vector-fields have disappeared
from the action \eqref{eq:model-action} altogether
and the cosmological constant problem remains unsolved.

Returning to the case of $\kappa\ne 0$ and $\zeta_{0}\ne 0$,
the above heuristic argument for obtaining a radiation-dominated
FRW universe, thus, relies on having an action employing the $Q_3$
field [which gives the $3\, H\, \dot{A}_0 + 3\, \dot{H}\,A_{0}$
combination in \eqref{eq:vector-field-eq-special}] and the
Ricci scalar (which vanishes precisely for $H=\half\,t^{-1}$).

An entirely different question is whether or not solution
\eqref{eq:A-H-aymptotic} appears dynamically as an attractor.
Here, this question will be addressed numerically,
leaving a proper mathematical treatment for the future
(see Note Added).

\subsection{Additional relativistic matter and dimensionless ODEs}
\label{subsec:Dimensionless-ODEs}

From now on, we also consider an additional standard-matter component
with energy density $\rho_M$ and pressure $P_M$.
These and other dimensional variables
can be replaced by the following dimensionless variables:
\bsubeqs\label{eq:dimensionless-variables}
\beqa
\big\{\Lambda,\, \epsilon,\,       \rho_M,\,       P_M    \big\}
&\to&
\big\{\lambda,\,           e,\, r_M,\, p_M \big\}\,,\\[2mm]
\big\{ t,  \,H,\,   Q_3,\, A_{0}  \big\}
&\to&
\big\{\tau,\,h,\,   q_3,\, v  \big\}\,,
\eeqa
\esubeqs
if appropriate powers of the (Planck-type) energy scale
$(8\pi\,G)^{-1/2}$ are used, without further numerical factors.
Henceforth, an overdot stands for differentiation with respect
to $\tau$, for example, $h(\tau)\equiv \dot{a}(\tau)/a(\tau)$.

The previous results \eqref{eq:Einstein-FRWeqs} and
\eqref{eq:vector-field-eq-special}, together with the extra
standard-matter component, give the following dimensionless ODEs:
\bsubeqs\label{eq:ODEs}
\beqa
\label{eq:ODEs-h}
&&
2\,\dot{h} + 3\, h^2 =
\lambda-
\zeta_{0}\,\big(\dot{v}+3\,  h\, v\big)^2-w_M\,r_M
\nonumber\\[2mm]&&
\hspace*{20mm}
+ 2\,\kappa\,\big[v^2\, \big( 4\,\dot{h}+9\,h^2\big)
      -2\,\dot{v}^2 -2\, v\, \ddot{v} -4\, h\, v\, \dot{v} \big]\,,
\\[2mm]\label{eq:ODEs-v}
&&
\ddot{v}+ 3\,h\,\dot{v}
+ \big[ 3\,\dot{h}+6\,(\kappa/\zeta_{0})(\dot{h}+2\,h^2)\,\big]\,v=0\,,
\\[2mm]\label{eq:ODEs-rM}
&&
\dot{r}_M + 3\,(1+w_M)\,h\,r_M = 0 \,,
\eeqa
\esubeqs
where the last equation describes the evolution of the
additional homogeneous perfect fluid of the
standard-matter component,
with constant EOS parameter $w_M\equiv p_M/r_M=1/3$
for ultrarelativistic particles.
The corresponding generalized Friedmann equation
from \eqref{eq:Friedmann-FRWeqs} is given by
\beqa\label{eq:Friedmann-ODE}
&&
3\, h^2 =\lambda-
\zeta_{0}\,\big(\dot{v}+3\,  h\, v\big)^2+r_M
-6\,\kappa\, \big(h^2\, v^2+2\,  h\,  v\, \dot{v}\big)\,.
\eeqa
Observe that the right-hand side of \eqref{eq:Friedmann-ODE},
evaluated for the asymptotic solution \eqref{eq:A-H-aymptotic},
has canceling $\text{O}(t^0)$ terms
for one particular value of $\overline{\theta}^{\,2}$, provided
condition \eqref{eq:zeta-kappa-condition-Ansatz} holds.
Specifically, the two possible $\overline{\theta}$ values are given by
\beqa\label{eq:theta-bar}
&&
\overline{\theta} =\pm\,
\sqrt{\frac{4\,\lambda/5}{6\,\kappa+5\,\zeta_{0}}}\,\;,
\eeqa
in terms of the dimensionless cosmological constant
$\lambda\equiv (8\pi\,G)^{2}\,\Lambda$. The actual sign of
$\overline{\theta}$ will be determined by the initial
boundary conditions.

In fact, the ODEs \eqref{eq:ODEs} are to be solved with
boundary conditions on
$v(\tau)$, $\dot{v}(\tau)$, $h(\tau)$, and $r_M(\tau)$
at $\tau=\tau_\text{start}$, where these particular
function values must satisfy
the constraint equation \eqref{eq:Friedmann-ODE}.
The corresponding physical quantities must be small enough
for classical gravity to be relevant,
e.g., $r_M(\tau_\text{start}) \ll 1$.

\vspace*{-0mm}

\subsection{Numerical results}
\label{subsec:Numerical-results}

Numerical calculations have been performed
for the model defined by \eqref{eq:model-action-Q3-general}
and \eqref{eq:f-epsilon-Ansaetze}, with the
dimensionless cosmological constant $\lambda=0.02$ and
model parameters $\kappa=-\zeta_{0}/2=-1/2$.
The results are presented in two figures, the first
without and the second with dynamical effects from the
vector-field.

\begin{figure*}[t] 
\begin{center}     
\hspace*{-4mm}
\includegraphics[width=1.05\textwidth]{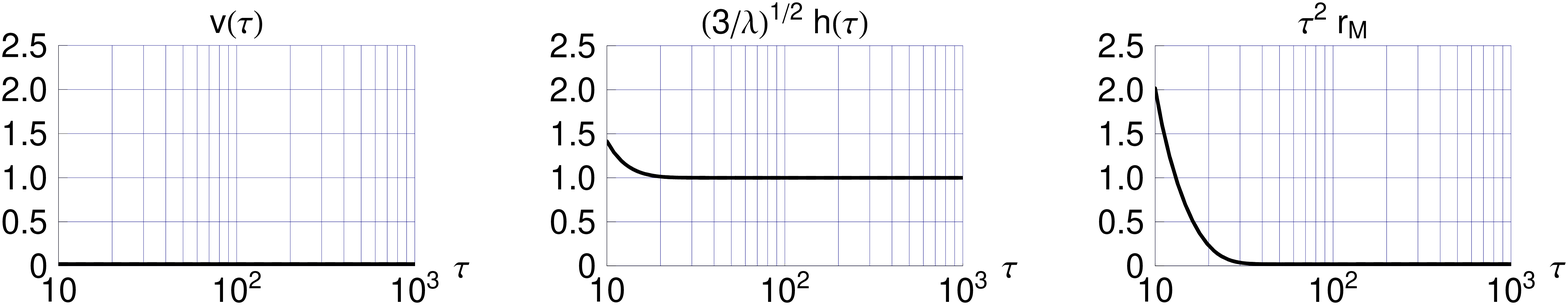}
\end{center}
\caption{Numerical solution of ODEs \eqref{eq:ODEs} for parameters
$\lambda = 0.02$, $\zeta_{0} = 1$, $\kappa=-1/2$, and $w_M=1/3$.
The boundary conditions are
$v(10)=\dot{v}(10)=0$, $r_M(10)=0.02$, and $h(10)=0.115470$.
With these boundary conditions, the ODEs have the
exact solution $v(\tau)=0$ for $\tau \geq 10$.}
\label{fig:1}
\begin{center}
\hspace*{-4mm}
\includegraphics[width=1.05\textwidth]{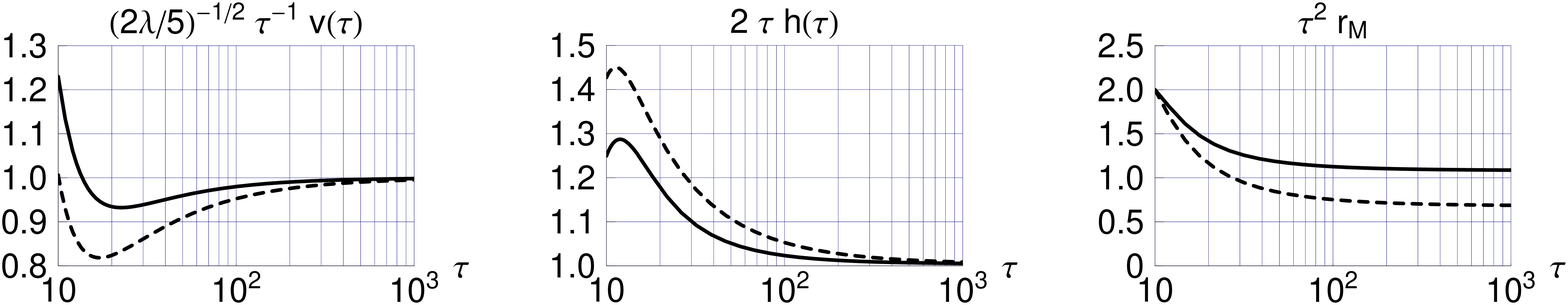}
\end{center}
\caption{Same as Fig.~\ref{fig:1}, but now with nonzero
starting values for the vector-field,
$v(10)$ $=$ $1 \pm 0.1$ and $\dot{v}(10)=0$,
where the dashed curves correspond to the starting value
$v(10)=0.9$. The corresponding starting values for $h$
follow from \eqref{eq:Friedmann-ODE} and are, respectively,
$h(10)=0.0624391$ and $h(10)=0.0713376$.
The scaling in the $v$--panel uses the
constant $\overline{\theta}$ from \eqref {eq:theta-bar},
which, for $\lambda = 0.02$, takes the value
$\overline{\theta}=\sqrt{2\,\lambda/5}=1/(5\,\sqrt{5})$.
The numerical solutions of $v(\tau)$ shown in the
left panel approach asymptotically the function
$\theta_\text{num}\,\tau$, with
$\theta_\text{num}=\overline{\theta}$ within the numerical
accuracy of the calculation
(the rescaled functions plotted in the left panel run towards
the value $1$ as  $\tau\to \infty$).
The corresponding numerical solutions of $h(\tau)$ shown in the
middle panel approach asymptotically the function
$\gamma_\text{num}\,\tau^{-1}$, with
$\gamma_\text{num}=1/2$ within the numerical
accuracy of the calculation
(the rescaled functions plotted in the middle panel run towards
the value $1$ as  $\tau\to \infty$).
}
\label{fig:2}
\vspace*{-0mm}
\end{figure*}

With special boundary conditions
to ensure the exact vanishing of the vector-field,
Fig.~\ref{fig:1} shows that
the cosmological constant $\lambda$ rapidly dominates
the matter component and that exponential expansion sets in.
Asymptotically, this model universe approaches
de Sitter space having $\dot{h}=r_M=0$.
The behavior of $h(\tau)$ and $r_M(\tau)$ shown in Fig.~\ref{fig:1}
is completely standard and can also be obtained analytically.

With boundary conditions of the vector-field in an
appropriate domain, Fig.~\ref{fig:2} shows that
the vector-field is able to compensate a Planck-scale
cosmological constant $\lambda=\text{O}(1)$
and that Minkowski spacetime is  approached asymptotically,
$h(\tau)\to 0$ for $\tau\to \infty$. The numerical solution
$v(\tau)$ asymptotically takes the form \eqref{eq:A-H-aymptotic},
with the precise linear coefficient needed for the complete
compensation of the cosmological constant $\lambda$.
It needs to be emphasized that this linear coefficient
of $v(\tau)$ is not put in by hand but arises
dynamically; cf. Ref.~\refcite{KV2010}.
See also the remark below \eqref{eq:Friedmann-ODE}
and the details given in the caption of Fig.~\ref{fig:2}.

The final phase shown in Fig.~\ref{fig:2} corresponds to an FRW
universe ($h \sim \half\,\tau^{-1}$) dominated by relativistic matter.
The solid curve in the right panel of Fig.~\ref{fig:2}
corresponds to an asymptotic universe with $r_M \sim 1.09\,\tau^{-2}$
and $r_V\equiv (3\,h^2 -r_M) \sim -0.34\,\tau^{-2}$
(hence, ratio $r_M/|r_V| \sim 3$)
and the dashed curve to an asymptotic universe with
$r_M \sim 0.687\,\tau^{-2}$ and $r_V\sim 0.063\,\tau^{-2}$
(hence, ratio $r_M/r_V \sim 11$).
For these numerical solutions, the action density term
$R\,A_{0}^2 \propto -6\,(\dot{h}+2\, h^2)\,v^2$ has also
been found to drop to zero faster than $\tau^{-1}$.

\newpage
\section{Discussion}
\label{sec:Discussion}

The vector-field model defined by \eqref{eq:model-action-Q3-general}
and \eqref{eq:f-epsilon-Ansaetze} provides an attractor solution
which compensates an arbitrary positive cosmological constant
$\Lambda$ and gives a final universe with
Hubble parameter $H(t) = \half\,t^{-1}$.
This final state resembles a standard
radiation-dominated FRW universe.\footnote{Most likely,
the model can be extended to allow for a similar compensation of
a cosmological constant $\Lambda$ of arbitrary
sign.\cite{KV2008a,KV2008b,KV2010,EmelyanovKlinkhamer2011-No1}
For the case of positive $\Lambda$, also a final cosmological phase with
$H(t) = (2/3)\,t^{-1}$ can be obtained
by replacing  $\epsilon(Q_3)$ in \eqref{eq:model-action}
with $\epsilon(Q_1,\,Q_2)=\zeta_1\,(Q_1)^2 + \zeta_2\,(Q_2)^2$,
for the contractions $(Q_1)^2 \equiv A_{\alpha;\beta}\,A^{\alpha;\beta}$
and $(Q_2)^2 \equiv A_{\alpha;\beta}\,A^{\beta;\alpha}$,
and by taking the coefficient $\kappa=-(\zeta_1 + \zeta_2)/2$
in \eqref{eq:f-Ansatz}.
This type of model may be relevant to
inflationary effects in vector-field theories with a dynamically
canceled cosmological constant.\cite{Klinkhamer2011}
At this point, it should be mentioned that certain scalar-tensor
theories (with a fundamental scalar field $\phi$) have also been argued to give
both a compensation of the cosmological constant and a final FRW-like
phase.\cite{Charmousis-etal2011}}

If the model is extended by
the addition of a standard noninteracting-relativistic-matter component,
the final state can be a genuine radiation-dominated FRW universe
with a subleading (time-dependent) vacuum-energy-density component.
Figure~\ref{fig:2} shows two possible model universes,
which start out with equal matter and vacuum energy
densities, $\rho_M=\Lambda \sim (E_\text{Planck})^4$,
but end up with the matter component dominating, $\rho_M>|\rho_V|$,
both components $\rho_M$ and $|\rho_V|$ decreasing as $t^{-2}$
(specific numbers are given in the last paragraph of
Sec.~\ref{subsec:Numerical-results}).

In this way, there is a more or less realistic physical description
of the earliest cosmological phase as a radiation-dominated FRW
universe with a dynamically canceled cosmological constant.
Possibly, this description needs to be augmented with
the effects from inflation.\cite{Klinkhamer2011}

As discussed in
Refs.~\refcite{KV2009-electroweak,K2010-electroweak,K2011-electroweak},
quantum-dissipative processes can be expected to lead to a
further (exponential) reduction of $|\rho_V(t)|$
in the very early universe ($k T \gg E_\text{ew} \sim \text{TeV}$)
and related processes
at the electroweak scale can perhaps generate a finite remnant
value of order $\rho_{V}(\infty) \sim (E_\text{ew}^2/E_\text{Planck})^4
\sim (\text{meV})^4$. Alternative explanations of the
remnant vacuum energy density in the $q$--theory framework
have been reviewed in Ref.~\refcite{KV2011-review}.

To conclude, both obstacles mentioned in the Introduction
have been dealt with separately, the first
in our previous article of Ref.~\refcite{EmelyanovKlinkhamer2011-No1}
and the second in the present article.
It remains to find a joint solution, provided such a solution exists.

\section*{Acknowledgments}

It is a pleasure to thank M. Kopp and G.E. Volovik for helpful discussions
and the referee for useful remarks.

\section*{Note Added}

For the case of $\lambda>0$ and $\kappa=-\zeta_{0}/2=-1/2$,
it is possible to prove the existence of an asymptotically
stable (attractor) solution. The proof relies on an
appropriate change of variables,
knowledge of the solution $r_M(a)$,
and the Poincar\'{e}--Lyapunov theorem
[Theorem 7.1 in Ref.~\refcite{Verhulst1996}].
Most likely, the attractor solution can be established rigorously
also for general  $\kappa\ne 0$ and $\zeta_{0}\ne 0$   
with $6\,\kappa+5\,\zeta_{0}>0$.

The present article is the second of a trilogy of articles,
the first one being Ref.~\refcite{EmelyanovKlinkhamer2011-No1}.
The third article of the trilogy
(Ref.~\refcite{EmelyanovKlinkhamer2011-No3})
finds the joint solution mentioned in the
last sentence of Sec.~\ref{sec:Discussion}.
This third article also gives a detailed mathematical discussion
of the attractor behavior in the type of vector-field models considered.

\end{document}